# Direct observation of double valence-band extrema and anisotropic effective masses of the thermoelectric material SnSe

Takanobu Nagayama[1], Kensei Terashima[2*], Takanori Wakita[2], Hirokazu Fujiwara[1], Tetsushi Fukura[1], Yuko Yano[1], Kanta Ono[3], Hiroshi Kumigashira[3], Osamu Ogiso[4,5], Aichi Yamashita[4,5], Yoshihiko Takano[4,5], Hitoshi Mori[6], Hidetomo Usui[6], Masayuki Ochi[6], Kazuhiko Kuroki[6], Yuji Muraoka[1,2], and Takayoshi Yokoya[1,2*]

[1]*Graduate School of Natural Science and Technology, Okayama University, Okayama 700-8530, Japan*

[2]*Research Institute for Interdisciplinary Science (RIIS), Okayama University, Okayama 700-8530, Japan*

[3]*Photon Factory, Institute of Materials Structure Science, High Energy Accelerator Research Organization (KEK), Tsukuba, Ibaraki 305-0801, Japan*

[4]*National Institute for Materials Science, Tsukuba, Ibaraki 305-0047, Japan*

[5]*Graduate School of Pure and Applied Science, University of Tsukuba, Tsukuba, Ibaraki 305-8577, Japan*

[6]*Department of Physics, Osaka University, Toyonaka, Osaka 560-0043, Japan*

*E-mail: k-terashima@cc.okayama-u.ac.jp; yokoya@cc.okayama-u.ac.jp

**ABSTRACT**

Synchrotron-based angle-resolved photoemission spectroscopy is used to determine the electronic structure of layered SnSe, which was recently turned out to be a potential thermoelectric material. We observe that the top of the valence band consists of two nearly independent hole bands, whose tops differ by ~20 meV in energy, indicating the necessity of a multivalley model to describe the thermoelectric properties. The estimated effective masses are anisotropic, with in-plane values of 0.16–0.39 $m_0$ and an out-of-plane value of 0.71 $m_0$, where $m_0$ is the rest electron mass. Information of the electronic structure is essential to further enhance the thermoelectric performance of hole-doped SnSe.

Thermoelectric materials facilitate direct conversion between a temperature difference and an electric current, and can be used to produce electric power from waste heat. Therefore, there has been extensive exploration of new thermoelectric materials.[1-3] Their performance is evaluated using the dimensionless figure of merit, $ZT = S^2\sigma T/\kappa$, where $S$, $\sigma$, $T$, and $\kappa$ are the Seebeck coefficient, electric conductivity, absolute temperature, and thermal conductivity, respectively. Higher $ZT$ leads to higher conversion efficiency.

Recently, it was reported that SnSe single crystals exhibit a high $ZT$ value of 2.6 along the $b$ axis [see Fig. 1(a) for the crystal structure] at $T$ = 923 K,[4] which exceeds the $ZT$ values of commonly used thermoelectric materials ($Bi_2Te_3$, PbTe, and $Si_{1-x}Ge_x$). The high $ZT$ value of SnSe is reported to originate from its extremely low thermal conductivity.[4,5] However, the power factor (PF = $S^2\sigma$), which determines the degree of the generation of voltage and current, is not so high. Therefore, optimizing the doping level, which dominantly controls the electronic part of the physical parameters, may improve PF and thus $ZT$. Very recently, intentionally hole-doped SnSe exhibited increased thermoelectric performance over a wide temperature range compared to non-doped SnSe.[6,7] Most importantly, the negligible $ZT$ values in the temperature region from 300 to 600 K in non-doped SnSe were markedly enhanced hole-doped SnSe. This may be most probably due to enhancement of the part of the PF by optimizing the doping level.

To understand the enhanced $ZT$ in this temperature region and to optimize doping level, knowledge of the electronic band structure is crucial. Electronic band calculations on SnSe have predicted that two valence-band extrema, or a "pudding-mold-like" shape, in the vicinity of the top of the valence band could be responsible for the improvement in the thermoelectric performance of hole-doped SnSe.[4,8-11] However, there seems to be no consensus on whether the shape of the top of the valence band is a single parabola[6] (with a degeneracy of 2 for SnSe) or multiple parabolas.[7] Regarding the electronic structure, the available experimental information on band dispersions of SnSe has been limited. Early photoemission studies reported only a momentum-integrated electronic structure.[12] Very recently, an angle-resolved photoemission spectroscopy (ARPES) study reported band dispersion, but detailed information on the effective mass has not yet been fully revealed.[13] Therefore, a more detailed study of the electronic structure is crucial and urgent in order to achieve higher thermoelectric performance in a SnSe system.

In this letter, we report on synchrotron-based ARPES experiments on single-crystalline SnSe, which are performed to experimentally determine the band dispersions of the top of the valence band. Synchrotron-based ARPES is capable of determining the electronic band structure over the whole momentum space.[14] We find that the top of the valence band consists of two nearly independent hole bands, indicating SnSe is a multivalley system. We determine their relative energies and anisotropic effective masses, and discuss the origin of the enhanced thermoelectric performance of hole-doped SnSe.

The SnSe single crystals were synthesized in the following two steps. First, stoichiometric amounts of Sn and Se were loaded into quartz tubes that were then evacuated and sealed. They were slowly heated to 600 °C and kept for 12 h, followed by furnace cooling. Second, the obtained compounds were ground and heated to 950 °C, and kept for 10 hours in the double-sealed evacuated quartz tubes. Single crystals were obtained after they were cooled down to 450 °C in 50 hours. The resistivity data of the samples exhibited a metallic behavior.

ARPES measurements were performed at BL-28A of the Photon Factory with circularly polarized light. All the data were obtained using a Scienta SES2002 electron analyzer, with the energy resolution set to 20 meV. The samples were cleaved *in situ* along the *a*-plane (parallel to the ΓZTY plane) at 20 K and under a base pressure of better than $1.7 \times 10^{-8}$ Pa. The samples were kept under the same conditions during measurement. The chemical potential $\mu$ of the samples was referenced to the Fermi level of a gold film that had a good electric contact with the samples. The lower measured temperatures helped to determine the energy positions of the bands with better accuracy compared to measurements made at higher temperatures, owing to smaller phonon broadening. No change of the spectral shape was observed during the measurement.

We performed a first-principles band structure calculation for SnSe in the Pnma phase using the WIEN2k package.[15] Experimental crystal structures at 295 K ($a$ = 11.501 Å, $b$ = 4.153 Å, and $c$ = 4.445 Å) were extracted from Ref. 16. The Perdew–Burke–Ernzerhof parametrization of the generalized gradient approximation[17] was used without including spin–orbit coupling because we verified that spin–orbit coupling has only a small effect on the band structures. We set $RK_{max}$ = 7.

Figures 1(b) and 1(c) show ARPES intensity maps of the valence bands of SnSe measured along $k_z$ in the bulk Brillouin zone (BZ) [Fig. 1(d)] with photon energies of 72 eV and 80 eV, respectively, together with the calculated band dispersions along the ΓZ and XU directions in the BZ (see the supplementary data 1). In both figures, the experimental band dispersions, which appear as bright regions, are dispersed symmetrically with respect to Γ(X). The top of the valence band is located very close to $\mu$ and slightly away from Z. Since the band gap of SnSe at room temperature is reported to be 0.86–0.898 eV,[4,18] the location of the valence band's top very close to $\mu$ indicates the *p*-type nature of the sample.

Several bands closely located between $\mu$ and 4 eV are seen, and these are mainly derived from Se 4p orbitals hybridized with Sn 5s,p orbitals. We also found separate bands with tops around Γ(X) and bottoms around Z(U) in the energy region from 5 to 8 eV, and these are dominant Sn 5s bands. We found an overall correspondence between the experimental and calculated band dispersions, indicating that band calculation can be a good guide to understanding the electronic structure of SnSe.

As the valence band data shows that $\mu$ is located very close to the top of the valence band, the electronic structure around the top of the valence band is closely related to the physical properties of SnSe. In order to experimentally confirm the location of the top of the valence band in the momentum space expected from the band calculations, we plot ARPES intensity maps around 0 and −300 meV with respect to $\mu$, respectively, in Figs. 2(a) and 2(b). In Fig. 2(a), there are two intensity spots on Γ–Z line: a higher intensity spot near Z and a lower intensity spot near the higher intensity spot. The momentum region of the higher intensity spot near Z is expanded in the −300-meV map [Fig. 2(b)], indicating that the observed spot near Z is the top of a hole band, which is consistent with the band calculations. The dispersion of a band corresponding to the lower intensity spot is not so clear, which will be discussed later. One can also notice a faint intensity spot on the ΓY line, which is the location of the top of the third-highest valence band, which is also consistent with band calculations.

We now focus on the band dispersions near $\mu$ in detail. Figures 2(c) and 2(d) show the ARPES intensity plot of SnSe and its minus second-derivative intensity plot, respectively, measured near the Z point and along the ΓZ line, where the top of the valence band is expected to be located by band calculations. The red lines in (d) are the calculated dispersions. In Fig. 2(c), one can observe several dispersive bands with tops near $\mu$. The uneven intensity may be due to the effects of matrix elements. In the second-derivative map, the dispersive bands symmetric with respect to Z can be seen more clearly. We determined the tops of two bands at approximately 0.46 (0.95) Å$^{-1}$ and 0.61 (0.80) Å$^{-1}$ (zone boundary 0.71 Å$^{-1}$), which we named as band 1 and band 2, respectively. The fact that $\mu$ is located above the top of the band dispersion, in spite of the metallic nature of the sample, may be due to the band bending expected from the doped semiconductors.[19]

Though the observed bands were similar to the results of the calculations, we found several discrepancies between the experimental and calculation results. One is the observation that the experimental data have a flat region at −150 and −450 meV at Z, as if the degeneration of the bands predicted from band calculations at Z is lifted in the experimental results. According to the band structure calculations, the two bands degenerate at −300 meV and at Z. Band structure calculations also predict that the two bands, as well as the other two bands that degenerate at −1.0 eV at Z, exhibit sizable dispersions along $k_x$. Since ARPES using vacuum ultraviolet light observes stationary points along $k_x$ because of the finite momentum window perpendicular to the measured surface along $k_x$ ($k_\perp$ broadening), the higher intensity regions around −150 meV and −450 meV may be related to bands along XU. Indeed, we can see a signature of a band dispersing to higher energy in Fig. 2(e) (orange arrow), which supports the absence of the lifting of degeneracy at Z. We also found that the bands forming the valence band maxima are two parabolic hole bands, and the expected hybridization gap of around 0.25 eV at $k_z$ = 0.57 (0.86) Å$^{-1}$ between those two bands to be negligibly small as compared to the calculation.[20]

Figure 2(e) shows a magnification of the dispersions near $\mu$ along ΓZ. It is more evident that the "pudding-mold" band predicted in the calculations,[9,11] which has been shown to drive large thermoelectric power in $Na_xCoO_2$,[21] is not clearly observed in the experiment. On the other hand, the observation of the two parabolic hole bands forming the valence band maxima indicates that a multivalley model is more suitable for discussing the enhanced low-temperature thermoelectric properties of hole-doped SnSe than a single-parabolic-band model.

Regarding the energy positions, the top of band 1 is closer to $\mu$ than that of band 2 by approximately 20 meV. The calculated energy difference between the maxima of the two bands may increase as a function of temperature,[9] and values of less than 1 meV (calculated from the measured lattice constants at 300 K)[8] and 60 meV (calculated from the measured lattice constants at 600 K and structure-relaxed)[7] have been reported (Table I), which is in line with the present study. The value calculated in this work is 32 meV, similar to the ARPES value. The difference in the values between the calculations may originate from the difference between the approximation and lattice parameters used. This work and Ref. 7 both performed density functional theory calculations with the same generalized gradient approximation but using different lattice parameters. This work uses lattice parameters from Ref. 16, while band structure of Ref. 7 is for a relaxed structure. Reference 8 reported the quasiparticle band dispersions calculated with GW approximation using the experimental lattice parameters from Ref. 22. The ARPES value of 20 meV at 20 K is consistent with the value of ~20 meV at 0 K deduced from temperature-dependent Hall coefficient measurements.[7]

We can deduce the effective mass along the $c$ and parallel to the $b$ directions on the measured momentum plane using a parabolic fit to the band positions, as shown in Figs. 2(e)–2(g). The deduced values are $m^*_c = 0.16\ m_0$ and $m^*_b = 0.38\ m_0$ for band 1 and $m^*_c = 0.17\ m_0$ and $m^*_b = 0.17\ m_0$ for band 2. These values are of the same order as the calculated values and are nearly the same as those reported by Zhao et al.[7] ($m^*_c = 0.14\ m_0$ and $m^*_b = 0.33\ m_0$ for band 1 and $m^*_c = 0.19\ m_0$ and $m^*_b = 0.18\ m_0$ for band 2), as compared in Table I. For band 1, we could observe the dispersion along the $k_x$ direction and estimate its effective mass $m^*_a$ to be ~0.7 $m_0$ [Fig. 2(h)]. In Table I, we use a value (0.71 $m_0$) determined from the data measured at 70 K, which is consistent with the value obtained from Fig. 2(h) but with better accuracy (see the supplementary data 2). Preliminary results showed that the $k_x$ dispersion of band 2 was ~4 times heavier than that of band 1.

It is known that larger numbers of valleys[2] and a higher anisotropy of effective mass at each valley (i.e., lower dimensionality of a system)[1,23,24] are favorable for increasing thermoelectric properties. While the former can be realized in a crystal with a higher symmetry, such as doped PbTe with a high-symmetry cubic structure,[25] the latter can be realized in a crystal with a lower symmetry. Direct observation of the characteristic electronic structures indicates that, in hole-doped SnSe with rather lower-symmetry structure, the unprecedented combination of the complicated band structure with multiple valleys and the anisotropy of effective mass from the layered structure yields enhanced thermoelectric performance in the lower-temperature region.

The present electronic structure information can be used for fine tuning of the thermoelectric properties by controlling the hole concentration. For higher-temperature regions, the valence band dispersions change as a result of the temperature-dependent variation of the lattice constants, even below the phase transition at 813 K.[9] The good correspondence of the band dispersions at 20 K between the experiment and calculations shown by the present work indicates that the band structure calculation is a good starting point for understanding the thermoelectric performance, even at higher temperatures.

Lastly, we would like to comment on the lifetime of carriers, which we may be able to deduce from the present ARPES results. Since the overlapping of several bands in SnSe and the large momentum intervals of the spectra for some measured directions have prevented us from estimating the lifetime for each band from the momentum distribution curves, we instead estimate the lifetime from the energy distribution curves (EDC) measured near the tops of the valence bands. The energy widths of the EDCs of band 1 and band 2 are 0.15–0.2 eV, which corresponds to a lifetime of around 3 fs. Note that this value is the shortest limit, because the width of the EDC includes the contribution from photoelectrons (Ref. 26) and other factors that would reduce the estimated lifetime.

In summary, we performed synchrotron-based ARPES experiments on SnSe single crystals to determine the band dispersions at the top of the valence band. The band structure was found experimentally to have two extrema of valence bands along the ΓZ line, with relatively smaller in-plane effective masses, which was consistent with band calculations. Direct observation of these characteristic electronic structures tells us that the multiple hole pockets and anisotropic effective masses play crucial roles in enhancing thermoelectric performance in the low-temperature region of hole-doped SnSe.

After completion of this work, we became aware of an ARPES study on SnSe,[27] which reported the in-plane effective masses for the two valence-band extrema and presented a similar discussion on the origin of the high thermoelectric performance.

Acknowledgments

H.F. is supported by a Grant-in-Aid for JSPS Fellows. This work was performed under the approval of the Photon Factory Program Advisory Committee (Proposal No. 2016G157). This work was partially supported by JSPS KAKENHI Grant Numbers 15H03691 and 16H04493 and by JST CREST (No. JPMJCR16Q6), Japan. This work was also partially supported by the Program for Promoting the Enhancement of Research Universities from MEXT.

**Figure Caption**

Fig. 1. (Color online) (a) Crystal structure of the low-temperature phase (*Pnma*) of SnSe.[28] SnSe exhibits a structural phase transition from the low-temperature phase to the high-temperature phase (*Cmcm*).[16] We choose the setting of the unit cell such that axis *a* is the longest and axis *b* is the shortest. ARPES intensity plots of the valence bands parallel to $k_z$ taken with photon energies of (b) 72 eV and (c) 80 eV, which are overlaid with the calculated bands along the ΓZ and XU lines. (d) BZ for the low-temperature crystal structure of SnSe. Hereafter, if not specified otherwise, calculated bands are expanded by a factor of 1.1 and shifted downward by 0.08 eV, and the data were taken with hν = 72 eV photons.

Fig. 2. (Color online) ARPES intensity maps integrated within ±50 at (a) 0 meV and (b) −300 meV with respect to *μ* measured on the ΓZTY plane. (c) ARPES intensity plot near *μ* along the ΓZ cut indicated on (a). (d) Minus second-derivative intensity plot of (c), together with calculated bands along ΓZ (red curves) and XU (red dotted curves). Minus second-derivative intensity plots near the top of the valence band, taken along (a) the ΓZ cut (but in the second BZ), (b) cut 2, and (c) cut 1. Note that the actual *k* locations of cut 1 and cut 2 are the second BZ, but the lines are placed at the corresponding *k* locations in the first BZ. The red circles in 2(e)–(g) are the band positions extracted from the EDCs by determining peak positions in the minus second-derivative intensities, while the blue curves are the results of fitting using parabolic functions. (h) Peak positions and results of fitting of band 1 perpendicular to the ΓZTY plane (see the supplementary data 2).

**Table I**

Comparisons of effective mass and relative energy of band 2 with respect to band 1 between present ARPES and band calculations (BC).

| | $m^*$ of Band 1 ($m_0$) | | | $m^*$ of Band 2 ($m_0$) | | | Relative energy (meV) |
|---|---|---|---|---|---|---|---|
| | $a$ | $b$ | $c$ | $a$ | $b$ | $c$ | |
| ARPES | 0.71 | 0.38 | 0.16 | 2–3 | 0.17 | 0.17 | 20 |
| BC (this work) | 0.52 | 0.33 | 0.14 | 1.20 | 0.14 | 0.16 | 32 |
| BC[7] | 0.76 | 0.33 | 0.14 | 2.49 | 0.18 | 0.19 | 60 |
| BC[8] | 0.74 | 0.16 | 0.31 | 0.90 | 0.15 | 0.12 | <1 |

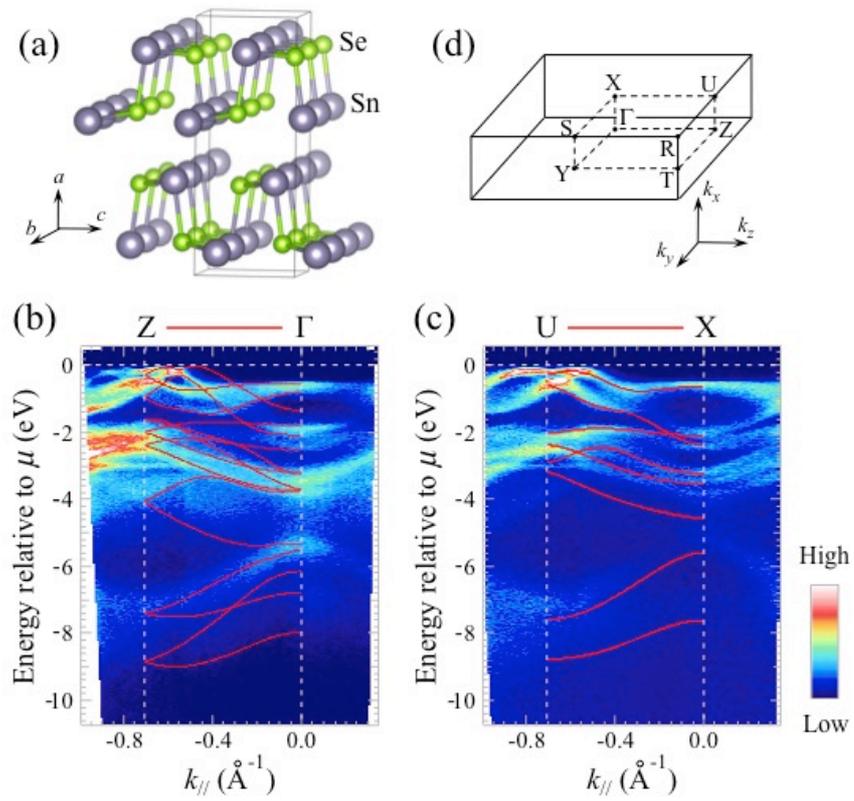

Fig. 1    Nagayama

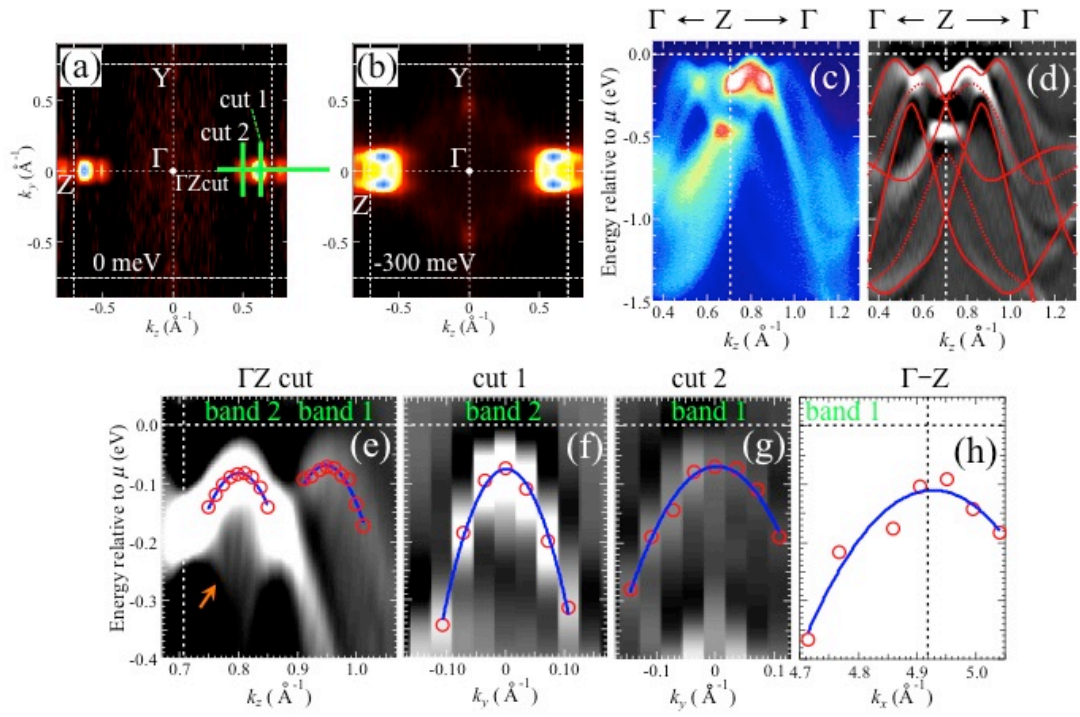

Fig. 2    Nagayama

**Supplementary data 1**

In an ARPES experiment, one can determine the momentum normal to the cleavage plane (*a* plane) by using a free-electron final-state model. However, $k_\perp$-broadening is expected not to be negligible, as, in the photon energy region used in the measurements (50–100 eV), the photoelectron escape depth is less than the length of axis *a* (the direction perpendicular to the measured surface). In order to obtain information on the measured $k_\perp$, we focused on particular bands that exhibit a sizable dispersion along this direction. In Supplemental data 1, we show the photon energy dependence of the minus second derivatives of the ARPES intensity maps measured along Γ(X)–Z(U) with photon energies from 56 eV to 96 eV. With varying photon energy, changes in the splitting and degeneracy of bands are seen, e.g., at $k_z \sim -0.6$ Å$^{-1}$ near *μ*, denoted with yellow squares. Among the photon energies used in this study, 72 eV shows the largest splitting of bands and 80 eV shows the highest degeneracy. Comparing the experimental bands with the calculated bands, the spectrum taken with 72 eV is similar to $k_\perp = 0$, as shown in Fig. 1(b), and that with 80 eV is similar to $k_\perp = \pi/a$, as shown in Fig. 1(c). Therefore, we consider that the ARPES spectra with photon energies of 72 eV and 80 eV are measured approximately on the ΓZTY and XURS planes, respectively.

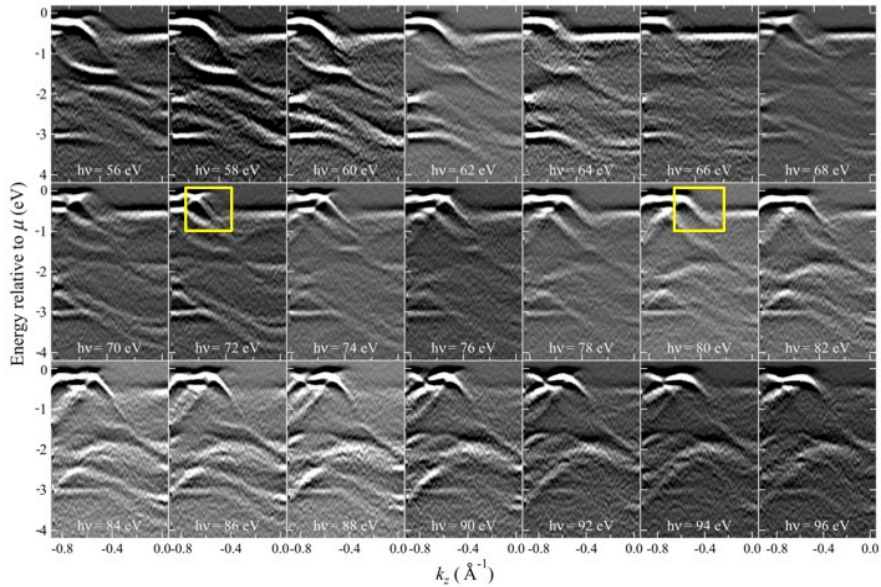

Supplementary data 1     Nagayama

**Supplementary data 2**

Minus second-derivative intensity plots near the top of the valence band at 70 K, taken perpendicular to the ΓZTY plane. The red circles are the band positions extracted from the EDCs by determining the peak positions in the minus second-derivative intensities, while the blue curve is the result of fitting using a parabolic function. The estimated effective mass $m^*_a$ is 0.71 $m_0$, which is consistent with that found at 20 K.

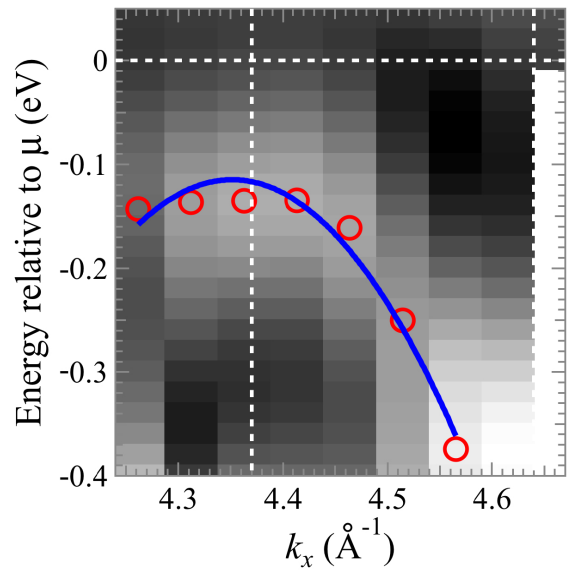

Supplementary data 2     Nagayama